\documentstyle[prd,aps,preprint,tighten,epsfig]{revtex}

\begin{document}

\title{Enhanced electromagnetic transition dipole moments and radiative decays
of massive Majorana neutrinos thanks to \\ the seesaw-induced
non-unitary effects}

\author{{\bf Zhi-zhong Xing} \thanks{E-mail: xingzz@ihep.ac.cn}
and {\bf Ye-Ling Zhou} \thanks{E-mail: zhouyeling@ihep.ac.cn}}
\address{Institute of High Energy Physics,
Chinese Academy of Sciences, Beijing 100049, China}

\maketitle

\begin{abstract}
In an extension of the standard electroweak theory where the
phenomenon of lepton flavor mixing is described by a $3\times 3$
unitary matrix $V$, the electric and magnetic dipole moments of
three active neutrinos are suppressed not only by their tiny masses
but also by the Glashow-Iliopoulos-Maiani (GIM) mechanism. We show
that it is possible to lift the GIM suppression if the canonical
seesaw mechanism of neutrino mass generation, which allows $V$ to be
slightly non-unitary, is taken into account. In view of current
experimental constraints on the non-unitarity of $V$, we find that
the effective electromagnetic transition dipole moments of three
light Majorana neutrinos and the rates of their radiative decays can
be maximally enhanced by a factor of ${\cal O}(10^2)$ and a factor
of ${\cal O}(10^4)$, respectively. This novel observation
reveals an intrinsic and presumably significant correlation between
the electromagnetic properties of massive neutrinos and the origin of
their small masses.
\end{abstract}


\newpage

\framebox{\bf 1} ~ The standard model (SM) of particle physics
contains three species of massless neutrinos which only take part in
weak interactions. Since 1998, a number of convincing neutrino
oscillation experiments have established the fact that neutrinos are
actually massive and lepton flavors mix in a way similar to the
quark flavor mixing \cite{PDG}. This breakthrough has lots of
profound and far-reaching impacts on cosmology and astrophysics, as
neutrinos are so abundant in the Universe that their tiny masses and
flavor conversions may significantly affect the cosmic expansion,
structure formation and many other astrophysical events \cite{XZ}.

The most popular mechanism of generating finite but tiny neutrino
masses beyond the SM is the canonical seesaw mechanism \cite{SS}, in
which the small neutrino masses are attributed to the existence of
heavy degrees of freedom such as the right-handed Majorana
neutrinos. In this elegant picture the $3\times 3$ lepton flavor
mixing matrix $V$ has a striking difference from the $3\times 3$
quark flavor mixing matrix in the SM: it is not exactly unitary due
to small mixing between light and heavy neutrinos as a result of the
Yukawa interactions. The deviation of $V$ from a unitary matrix can
be appreciable, in particular when the seesaw mechanism works at the
TeV scale so as to avoid the seesaw-induced hierarchy problem
\cite{Vissani} and satisfy the testability requirement at the Large
Hadron Collider \cite{Xing09}. A careful analysis of current data on
the invisible width of the $Z^0$ boson, universality tests of
electroweak interactions, rare leptonic decays and neutrino
oscillations indicates that the unitarity of $V$ is good at the
percent level and possible non-unitary effects are also allowed at
the same level \cite{Antusch}.

Now that three active neutrinos acquire their respective masses in
the seesaw mechanism, they should have the electromagnetic dipole
moments (EMDMs) through quantum loops. The fact that the Majorana
neutrinos are their own antiparticles implies that they can only
have the {\it transition} EMDMs between two different neutrino mass
eigenstates in an electric or magnetic field \cite{XZ}. The relevant
radiative decays of the heavier active neutrinos, which may
contribute to the cosmic infrared background in the Universe
\cite{CIB}, are of particular interest in cosmology. Since all the
previous calculations of the EMDMs and radiative decays of massive
neutrinos were done by assuming $V$ to be exactly unitary, we find
it highly necessary and important to recalculate the same quantities
by taking into account the seesaw-induced non-unitary effects.

The aim of this paper is just to examine the seesaw-induced
non-unitary effects on the EMDMs and radiative decays of active
Majorana neutrinos. We do a complete one-loop calculation and
demonstrate that such effects are likely to be significantly larger
than the standard (unitary) contributions, because the suppression
induced by the Glashow-Iliopoulos-Maiani (GIM) mechanism \cite{GIM}
in the latter case can now be lifted. A careful numerical analysis
shows that the effective EMDMs of three neutrinos and the rates of
their radiative decays can be maximally enhanced by a factor of
${\cal O}(10^2)$ and a factor of ${\cal O}(10^4)$. Such an intrinsic
and presumably important correlation between the electromagnetic
properties of massive neutrinos and the origin of their small masses
must be taken seriously, and it may even serve as a sensitive
touch-stone for the highly-regarded seesaw mechanism in the future.

\vspace{0.4cm}

\framebox{\bf 2} ~ The canonical seesaw mechanism is based on a
simple extension of the SM in which three heavy right-handed
neutrinos are added and the lepton number is violated by their
Majorana mass term \cite{SS}:
\begin{eqnarray}
-{\cal L}_\nu = \overline{\ell^{}_{\rm L}} Y^{}_\nu \tilde{H}
N^{}_{\rm R} + \frac{1}{2} \overline{N^c_{\rm R}} M^{}_{\rm R}
N^{}_{\rm R} + {\rm h.c.} \; ,
\end{eqnarray}
where $\tilde{H} \equiv i \sigma^{}_2 H^*$ with $H$ being the SM
Higgs doublet, $\ell^{}_{\rm L}$ denotes the left-handed lepton
doublet, $N^{}_{\rm R}$ stands for the column vector of three
right-handed neutrinos, and $M^{}_{\rm R}$ is a symmetric Majorana
mass matrix. After spontaneous $SU(2)^{}_{\rm L} \otimes
U(1)^{}_{\rm Y} \to U(1)^{}_{\rm em}$ gauge symmetry breaking, $H$
achieves its vacuum expectation value $\langle H\rangle =
v/\sqrt{2}$ with $v \simeq 246$ GeV. Then the Yukawa-interaction
term in ${\cal L}^{}_\nu$ yields the Dirac mass matrix $M^{}_{\rm D}
= Y^{}_\nu v/\sqrt{2}$, but the Majorana mass term in ${\cal
L}^{}_\nu$ keeps unchanged because right-handed neutrinos are the
$SU(2)^{}_{\rm L}$ singlet and thus they are not subject to the
electroweak symmetry breaking. The overall neutrino mass matrix
turns out to be a symmetric $6\times 6$ matrix and can be
diagonalized through
\begin{eqnarray}
{\cal U}^\dagger \left( \matrix{ {\bf 0} & M^{}_{\rm D} \cr M^T_{\rm
D} & M^{}_{\rm R}}\right) {\cal U}^* = \left( \matrix{
\widehat{M}^{}_\nu & {\bf 0} \cr {\bf 0} & \widehat{M}^{}_N}\right)
\; ,
\end{eqnarray}
in which we have defined $\widehat{M}^{}_\nu \equiv {\rm
Diag}\{m^{}_1, m^{}_2, m^{}_3 \}$ and $\widehat{M}^{}_N \equiv {\rm
Diag}\{M^{}_1, M^{}_2, M^{}_3 \}$ with $m^{}_i$ and $M^{}_i$ being
the physical masses of three light neutrinos $\nu^{}_i$ and three
heavy neutrinos $N^{}_i$ (for $i=1,2,3$). The $6\times 6$ unitary
matrix $\cal U$ is decomposed as \cite{Xing11}
\begin{eqnarray}
{\cal U} = \left( \matrix{ {\bf 1} & {\bf 0} \cr {\bf 0} & U^{}_0
\cr} \right) \left( \matrix{A & R \cr S & B \cr} \right) \left(
\matrix{V^{}_0 & {\bf 0} \cr {\bf 0} & {\bf 1} \cr} \right) \; ,
\end{eqnarray}
where ${\bf 1}$ denotes the $3\times 3$ identity matrix, $U^{}_0$
and $V^{}_0$ are the $3\times 3$ unitary matrices, and $A$, $B$, $R$
and $S$ are the $3\times 3$ matrices which characterize the
correlation between the active or light neutrino sector ($V^{}_0$)
and the sterile or heavy neutrino sector ($U^{}_0$). A full
parametrization of $\cal U$ in terms of 15 mixing angles and 15
CP-violating phases has been given in Ref. \cite{Xing11}. One may
express the flavor eigenstates of three active neutrinos in terms of
the mass eigenstates $\nu^{}_i$ and $N^{}_i$. In the mass basis of
charged leptons and neutrinos, the leptonic weak charged-current
(cc) and neutral-current (nc) interactions read
\begin{eqnarray}
-{\cal L}^{}_{\rm cc} & = & \frac{g}{\sqrt{2}} \left[
\overline{l^{}_{\alpha \rm L}} \gamma^\mu \left( V^{}_{\alpha i}
\nu^{}_{i \rm L} + R^{}_{\alpha i} N^{}_{i
\rm L} \right) W^{-}_\mu + {\rm h.c.} \right] \; , \nonumber \\
-{\cal L}^{}_{\rm nc} & = & \frac{g}{2\cos\theta^{}_{\rm w}} \left\{
\overline{\nu^{}_{i \rm L}} \gamma^\mu (V^\dagger V)^{}_{ij}
\nu^{}_{j \rm L} + \overline{N^{}_{i \rm L}} \gamma^\mu (R^\dagger
R)^{}_{ij} N^{}_{j \rm L} + \left[\overline{\nu^{}_{i \rm L}}
\gamma^\mu (V^\dagger R)^{}_{ij} N^{}_{j \rm L} + {\rm h.c.} \right]
\right\} Z^{}_\mu \; ,
\end{eqnarray}
where $\alpha$ runs over $e$, $\mu$ or $\tau$, $V = AV^{}_0$ is
responsible for the flavor mixing of active neutrinos $\nu^{}_i$,
and $R$ measures the strength of charged-current interactions of
heavy neutrinos $N^{}_i$ (for $i=1,2,3$) \cite{Xing08}. A small
deviation of $V$ from $V^{}_0$ is actually characterized by
nonvanishing $R$, as $VV^\dagger = AA^\dagger = {\bf 1} -
RR^\dagger$ holds. The exact seesaw relation between the masses of
light and heavy neutrinos is $V \widehat{M}^{}_\nu V^T + R
\widehat{M}^{}_N R^T = {\bf 0}$, which signifies the correlation
between neutrino masses and flavor mixing parameters.

Let us consider the radiative $\nu^{}_i \to \nu^{}_j + \gamma$
transition, whose electromagnetic vertex can be written as
\begin{eqnarray}
\Gamma^\mu_{ij}(0) = \mu^{}_{ij} \left(i \ \sigma^{\mu\nu}
q^{}_\nu\right) + \epsilon^{}_{ij} \left(\sigma^{\mu\nu} q^{}_\nu
\gamma^{}_5 \right) \;
\end{eqnarray}
for a real photon satisfying the on-shell conditions $q^2 =0$ and
$q^{}_\mu \varepsilon^\mu =0$. In Eq. (5) $\epsilon^{}_{ij}$ and
$\mu^{}_{ij}$ are the electric and magnetic {\it transition} dipole
moments of Majorana neutrinos, and their sizes can be calculated via
the proper vertex diagrams in FIG. 1 (weak cc interactions). The
$\gamma$-$Z$ self-energy diagrams in FIG. 2 (weak nc interactions)
do not have any {\it net} contribution to $\epsilon^{}_{ij}$ and
$\mu^{}_{ij}$, but we find that they play a very crucial role in
eliminating the infinities because the divergent terms originating
from FIG. 1 are unable to automatically cancel out in the presence
of the seesaw-induced non-unitary effects (i.e., $R \neq {\bf 0}$
and $V \neq V^{}_0$) unless those divergent terms originating from
FIG. 2 are also taken into account. This observation is new. It
implies that the non-unitary case under discussion is somewhat
different from the unitary case (i.e., $R={\bf 0}$ and $V^\dagger V
= V^\dagger_0 V^{}_0 ={\bf 1}$) discussed before in the literature
\cite{Shrock}, where the Feynman diagrams in FIG. 2 are forbidden
and the divergent terms arising from FIG. 1 can automatically cancel
out.

After a careful treatment of the infinities and non-unitary effects
in our calculations \cite{XZ12}, we arrive at
\begin{eqnarray}
\mu^{}_{ij} & = & \frac{i e G^{}_{\rm F}}{4\sqrt{2} \pi^2}
\left(m^{}_i + m^{}_j \right) \sum^{}_{\alpha} F^{}_\alpha \text{Im}
\left(V^{}_{\alpha i} V^*_{\alpha j} \right) \; ,
\nonumber \\
\epsilon^{}_{ij} & = & \frac{e G^{}_{\rm F}}{4\sqrt{2} \pi^2}
\left(m^{}_i - m^{}_j \right) \sum^{}_{\alpha} F^{}_\alpha \text{Re}
\left(V^{}_{\alpha i} V^*_{\alpha j} \right) \; ,
\end{eqnarray}
where
\begin{eqnarray}
F^{}_\alpha = \frac{3}{4} \left[
\frac{2-\xi^{}_\alpha}{1-\xi^{}_\alpha} - \frac{2
\xi^{}_\alpha}{\left(1-\xi^{}_\alpha\right)^2} + \frac{2
\xi^2_\alpha\ln \xi^{}_\alpha}{\left(1-\xi^{}_\alpha\right)^3}
\right] \;
\end{eqnarray}
with $\xi^{}_\alpha \equiv m^2_\alpha/M^2_W$ (for $\alpha = e, \mu,
\tau$) denotes the one-loop function. Although this result is {\it
formally} the same as that obtained in Ref. \cite{Shrock}, they are
{\it intrinsically} different as the seesaw-induced non-unitary
effects on $\mu^{}_{ij}$ and $\epsilon^{}_{ij}$ were not considered
in the previous works. To see how important such non-unitary effects
may be, let us make two analytical approximations. First,
$F^{}_\alpha \simeq 3 \left(2- \xi^{}_\alpha \right)/4$ holds to a
good degree of accuracy for $\xi^{}_\alpha \ll 1$. Second, $V =
AV^{}_0 \simeq V^{}_0 - T V^{}_0$ is also a good approximation for
small non-unitary corrections to $V^{}_0$, where \cite{Xing11}
\begin{eqnarray}
V^{}_0 & = & \left( \matrix{ c^{}_{12} c^{}_{13} & \hat{s}^*_{12}
c^{}_{13} & \hat{s}^*_{13} \cr -\hat{s}^{}_{12} c^{}_{23} -
c^{}_{12} \hat{s}^{}_{13} \hat{s}^*_{23} & c^{}_{12} c^{}_{23} -
\hat{s}^*_{12} \hat{s}^{}_{13} \hat{s}^*_{23} & c^{}_{13}
\hat{s}^*_{23} \cr \hat{s}^{}_{12} \hat{s}^{}_{23} - c^{}_{12}
\hat{s}^{}_{13} c^{}_{23} & -c^{}_{12} \hat{s}^{}_{23} -
\hat{s}^*_{12} \hat{s}^{}_{13} c^{}_{23} & c^{}_{13} c^{}_{23} \cr}
\right) ,
\nonumber \\
T & = & \left( \matrix{ \displaystyle\frac{1}{2} \sum^6_{k=4}
s^2_{1k} & 0 & 0 \cr \displaystyle\sum^6_{k=4} \hat{s}^{}_{1k}
\hat{s}^*_{2k} & \displaystyle\frac{1}{2} \sum^6_{k=4} s^2_{2k} & 0
\cr \displaystyle\sum^6_{k=4} \hat{s}^{}_{1k} \hat{s}^*_{3k} &
\displaystyle\sum^6_{k=4} \hat{s}^{}_{2k} \hat{s}^*_{3k} &
\displaystyle\frac{1}{2} \sum^6_{k=4} s^2_{3k} \cr} \right) ~
\end{eqnarray}
with $c^{}_{ij} \equiv \cos\theta^{}_{ij}$ and $\hat{s}^{}_{ij}
\equiv e^{i\delta^{}_{ij}} \sin\theta^{}_{ij}$ (here
$\theta^{}_{ij}$ and $\delta^{}_{ij}$ are the mixing angles and
CP-violating phases). Note that the light-heavy neutrino mixing
angles $\theta^{}_{ik}$ (for $i=1,2,3$ and $k=4,5,6$) are at most of
${\cal O}(0.1)$ \cite{Antusch}, such that the deviation of $V$ from
$V^{}_0$ is at the percent level or much smaller. Then we obtain
\begin{eqnarray}
\sum_\alpha F^{}_\alpha \left(V^{}_{\alpha i} V^*_{\alpha j} \right)
\simeq -\frac{3}{2} \sum_\alpha \left[\left(V^{}_0\right)^{}_{\alpha
i} \left(T V^{}_0\right)^*_{\alpha j} + \left(T
V^{}_0\right)^{}_{\alpha i} \left(V^{}_0\right)^*_{\alpha j} \right]
- \frac{3}{4} \sum_\alpha \left[ \xi^{}_\alpha
\left(V^{}_0\right)^{}_{\alpha i} \left(V^{}_0\right)^*_{\alpha j}
\right] \; .
\end{eqnarray}
The first and second terms on the right-hand side of this equation
correspond to the non-unitary and unitary contributions,
respectively. While the former is suppressed by $s^2_{ik} \lesssim
{\cal O}(10^{-2})$ (for $i=1,2,3$ and $k=4,5,6$) hidden in $T$, the
latter is suppressed by $\xi^{}_\alpha \lesssim 4.9 \times 10^{-4}$
(for $\alpha =e, \mu, \tau$) due to the GIM mechanism. We therefore
draw a generic conclusion that the seesaw-induced non-unitary
effects on $\epsilon^{}_{ij}$ and $\mu^{}_{ij}$ can be comparable
with or even larger than the standard (unitary) contributions.

The rates of radiative $\nu^{}_i \to \nu^{}_j + \gamma$ decays are
more sensitive to the non-unitarity of $V$, since they directly
depend on $|\mu^{}_{ij}|^2$ and $|\epsilon^{}_{ij}|^2$. Namely,
\begin{eqnarray}
\Gamma^{}_{\nu^{}_i \to \nu^{}_j + \gamma} = \frac{\left(
m^2_i-m^2_j \right)^3}{8\pi m^3_i} \left(|\mu^{}_{ij}|^2_{} +
|\epsilon^{}_{ij}|^2_{} \right) \simeq 5.3 \times \left(1 -
\frac{m^2_j}{m^2_i} \right)^3 \left(\frac{m^{}_i}{{\rm 1 ~
eV}}\right)^3 \left(\frac{\mu^{}_{\rm eff}}{\mu^{}_{\rm B}}\right)^2
{\rm s}^{-1}
\end{eqnarray}
with $\mu^{}_{\rm eff} \equiv \sqrt{|\mu^{}_{ij}|^2 +
|\epsilon^{}_{ij}|^2}$ for $\nu^{}_i\to \nu^{}_j + \gamma$ being the
effective EMDMs and $\mu^{}_{\rm B} = e/(2m^{}_e)$ being the Bohr
magneton. The size of $\Gamma^{}_{\nu^{}_i \to \nu^{}_j + \gamma}$
can be experimentally constrained by observing no emission of the
photons from solar $\nu^{}_e$ and reactor $\overline{\nu}^{}_e$
fluxes. More stringent constraints on $\mu^{}_{\rm eff}$ come from
the Supernova 1987A limit on neutrino decays and from the
cosmological limit on distortions of the cosmic microwave background
radiation (in particular, its infrared part): $\mu^{}_{\rm eff} <
{\rm a ~ few} \times 10^{-11} ~\mu^{}_{\rm B}$ \cite{Raffelt99}. Now
that more and more interest is being paid to the cosmic infrared
background relevant to the radiative decays of massive neutrinos
\cite{CIB}, it is desirable to evaluate $\mu^{}_{\rm eff}$ and
$\Gamma^{}_{\nu^{}_i \to \nu^{}_j + \gamma}$ on a well-defined
theoretical ground, such as the canonical seesaw mechanism under
discussion.

\vspace{0.4cm}

\framebox{\bf 3} ~ We proceed to numerically illustrate the
non-unitary effects on $\mu^{}_{\rm eff}$ and
$\Gamma^{}_{\nu^{}_i\to \nu^{}_j + \gamma}$. In view of current
neutrino oscillation data \cite{Fogli}, we take $\Delta m^2_{21}
\simeq 7.6 \times 10^{-5} ~{\rm eV}^2$, $\Delta m^2_{32} \simeq \pm
2.5 \times 10^{-3} ~{\rm eV}^2$, $\theta^{}_{12} \simeq 34^\circ$,
$\theta^{}_{23} \simeq 45^\circ$ and $\theta^{}_{13} \simeq 9^\circ$
\cite{DYB} as our typical inputs. The mass scale of three neutrinos,
the sign of $\Delta m^2_{32}$ and the values of three CP-violating
phases of $V^{}_0$ remain unknown. In our numerical calculation we
consider both normal ($\Delta m^2_{32} >0$) and inverted ($\Delta
m^2_{32} <0$) hierarchies by fixing the smallest neutrino mass to be
5 meV, and allow all the CP-violating phases to vary between zero
and $2\pi$. Those small active-sterile neutrino mixing angles in Eq.
(8) are constrained by present experimental data \cite{Antusch} as
follows:
\begin{eqnarray}
&& T^{}_{11} < 5.5 \times 10^{-3} \; , ~~~~ \left| T^{}_{21} \right|
< 7.0 \times 10^{-5} \; ,
\nonumber \\
&& T^{}_{22} < 5.0 \times 10^{-3} \; , ~~~~ \left| T^{}_{31} \right|
< 1.6 \times 10^{-2} \; ,
\nonumber \\
&& T^{}_{33} < 5.0 \times 10^{-3} \; , ~~~~ \left| T^{}_{32} \right|
< 1.0 \times 10^{-2} \; .
\end{eqnarray}
Note that all $s^{}_{ik}$ in $T$ (for $i=1,2,3$ and $k=4,5,6$) are
positive or vanishing. The CP-violating phases $\delta^{}_{ik}$ are
all allowed to vary from zero to $2\pi$, but they must satisfy the
above constraints together with the relations $VV^\dagger +
RR^\dagger = {\bf 1}$ and $V \widehat{M}^{}_\nu V^T + R
\widehat{M}^{}_N R^T = {\bf 0}$. It is in principle possible to
determine the masses of three heavy Majorana neutrinos with the help
of the exact seesaw relation, if all the other parameters are known
\cite{Xing2009}. Although such a prediction is in practice
impossible, we find that the upper bound on the non-unitarity of $V$
implies the existence of a lower bound on $M^{}_i$. In other words,
the values of $M^{}_i$ cannot be too small (i.e., they should be far
away from those of $m^{}_i$) so as to avoid too significant
non-unitary effects. To assure that radiative corrections to the
masses of three light neutrinos (via the one-loop self-energy
diagrams involving the heavy neutrinos) are sufficiently small
(e.g., smaller than $0.5$ meV) and stable, we simply assume that the
masses of three heavy neutrinos are nearly degenerate
\cite{Pilaftsis} and not more than ${\cal O}(1)$ TeV. This
assumption implies that we are concentrating on a limited and safe
parameter space of the TeV seesaw mechanism, but it is instructive
enough to reveal the salient features of the non-unitary effects on
the effective EMDMs $\mu^{}_{\rm eff} (\nu^{}_i\to \nu^{}_j +
\gamma)$ and the radiative decay rates $\Gamma^{}_{\nu^{}_i\to
\nu^{}_j + \gamma}$.

To present our numerical results in a convenient way, let us define
\begin{eqnarray}
\varepsilon^{}_{\rm uv} \equiv \left[\sum^6_{k=4} \left( s^2_{1k} +
s^2_{2k} + s^2_{3k} \right)\right]^{1/2} \; ,
\end{eqnarray}
which measures the overall strength of the unitarity violation of
$V$, and $\varepsilon^{}_{\rm uv} \in [0, 0.15)$ is reasonably taken
in our calculations. Namely, we allow each $s^{}_{ik}$ (for
$i=1,2,3$ and $k=4,5,6$) to vary in the range $0 \leq s^{}_{ik} <
0.15$. The numerical dependence of $\mu^{}_{\rm eff} (\nu^{}_i\to
\nu^{}_j + \gamma)$ and $\Gamma^{}_{\nu^{}_i \to \nu^{}_j + \gamma}$
on $\varepsilon^{}_{\rm uv}$ is shown in FIGs. 3 and 4,
respectively. Some discussions are in order.

(1) Switching off the non-unitary effects (i.e.,
$\varepsilon^{}_{\rm uv} =0$), we obtain the effective
electromagnetic dipole moments
\begin{eqnarray}
\mu^{}_{\rm eff} \simeq \left\{ \begin{array}{l} \left(0.8 \sim
3.0\right) \times 10^{-25} ~\mu^{}_{\rm B} ~~~~~~ (\nu^{}_2 \to
\nu^{}_1 + \gamma) \; , \\
\left(0.8 \sim 1.5\right) \times 10^{-24} ~\mu^{}_{\rm B} ~~~~~~
(\nu^{}_3 \to \nu^{}_1 + \gamma) \; , \\
\left(1.1 \sim 2.1\right) \times 10^{-24} ~\mu^{}_{\rm B} ~~~~~~
(\nu^{}_3 \to \nu^{}_2 + \gamma) \; , \end{array} \right .
\end{eqnarray}
for the normal mass hierarchy with $m^{}_1 \simeq 5$ meV; and
\begin{eqnarray}
\mu^{}_{\rm eff} \simeq \left\{ \begin{array}{l} \left(0.01 \sim 2.0
\right) \times 10^{-24} ~\mu^{}_{\rm B} ~~~~~ (\nu^{}_2
\to \nu^{}_1 + \gamma) \; , \\
\left(0.8 \sim 1.5\right) \times 10^{-24} ~\mu^{}_{\rm B} ~~~~~~\;
(\nu^{}_3 \to \nu^{}_1
+ \gamma) \; , \\
\left(1.3 \sim 2.0\right) \times 10^{-24} ~\mu^{}_{\rm B} ~~~~~~\;
(\nu^{}_3 \to \nu^{}_2 + \gamma) \; , \end{array} \right .
\end{eqnarray}
for the inverted mass hierarchy with $m^{}_3 \simeq 5$ meV, where
the uncertainties mainly come from the unknown CP-violating phases
$\delta^{}_{12}$, $\delta^{}_{13}$ and $\delta^{}_{23}$. Such
standard (unitary) results are far below the observational upper
bound on $\mu^{}_{\rm eff}$ ($<$ a few $\times 10^{-11} ~\mu^{}_{\rm
B}$ \cite{Raffelt99}), but they serve as a good reference to the
non-unitary effects on $\mu^{}_{\rm eff}$ being explored in this
work.

(2) FIGs. 3 and 4 clearly show that $\mu^{}_{\rm eff}$ and
$\Gamma^{}_{\nu^{}_i \to \nu^{}_j + \gamma}$ can be maximally
enhanced by a factor of ${\cal O}(10^2)$ and a factor of ${\cal
O}(10^4)$, respectively, in particular when $\varepsilon^{}_{\rm
uv}$ approaches its upper limit as set by current experimental data.
Note that the magnitude of $\mu^{}_{\rm eff} (\nu^{}_2 \to \nu^{}_1
+ \gamma)$ may be strongly suppressed in the inverted neutrino mass
hierarchy. The reason is rather simple: on the one hand, $m^{}_1
\simeq m^{}_2$ holds in this case, and thus $\epsilon^{}_{12}
\propto (m^{}_2 - m^{}_1)$ must be very small; on the other hand,
$\mu^{}_{12}$ depends on ${\rm Im}(V^{}_{\alpha 1} V^*_{\alpha 2})$,
so it can also be very small when the CP-violating phases are around
zero or $\pi$. This two-fold suppression becomes severer for the
decay rate $\Gamma^{}_{\nu^{}_2 \to \nu^{}_1 + \gamma}$, because it
is proportional to $ (m^{}_2 - m^{}_1)^3 \mu^{2}_{\rm eff} (\nu^{}_2
\to \nu^{}_1 + \gamma)$.

(3) Our numerical analysis shows that the results of $\mu^{}_{\rm
eff}$ and $\Gamma^{}_{\nu^{}_i \to \nu^{}_j + \gamma}$ are sensitive
to the absolute neutrino mass scale for both normal and inverted
mass hierarchies. For instance, $\mu^{}_{\rm eff} (\nu^{}_2 \to
\nu^{}_1 + \gamma)$ and $\mu^{}_{\rm eff} (\nu^{}_3 \to \nu^{}_1 +
\gamma)$ get enhanced when $m^{}_1$ changes from zero to 5 meV in
the normal mass hierarchy; while $\mu^{}_{\rm eff} (\nu^{}_1 \to
\nu^{}_3 + \gamma)$ and $\mu^{}_{\rm eff} (\nu^{}_2 \to \nu^{}_3 +
\gamma)$ are enhanced when $m^{}_3$ changes from zero to 5 meV in
the inverted mass hierarchy. This kind of sensitivity is not so
obvious if one only takes a look at the expressions of $\mu^{}_{ij}$
and $\epsilon^{}_{ij}$ in Eq. (6). The main reason is that a change
of $m^{}_1$ or $m^{}_3$ requires some fine-tuning of the
active-sterile neutrino mixing angles and CP-violating phases as
dictated by the exact seesaw relation $V \widehat{M}^{}_\nu V^T + R
\widehat{M}^{}_N R^T = {\bf 0}$, leading to a possibly significant
change of $\mu^{}_{\rm eff}$. The dependence of $\Gamma^{}_{\nu^{}_i
\to \nu^{}_j + \gamma}$ on the absolute neutrino mass scale is
somewhat more complicated, as one can see from Eq. (10).

(4) We find that the CP-violating phases play a very important role
in fitting both the exact seesaw relation and Eq. (11). If the heavy
neutrino masses $M^{}_i$ are not suppressed, then an appreciable
value of $\varepsilon^{}_{\rm uv}$ requires some fine cancellations
in the matrix product $R \widehat{M}^{}_N R^T$ such that
sufficiently small $m^{}_i$ can be obtained from $V
\widehat{M}^{}_\nu V^T = -R \widehat{M}^{}_N R^T$.
On the other hand, we remark that it is actually unnecessary
to require $M^{}_i$ to be around or above the electroweak
scale. The seesaw-induced non-unitary effects on $\mu^{}_{\rm eff}$
and $\Gamma^{}_{\nu^{}_i \to \nu^{}_j + \gamma}$ can be significant
even if one allows one, two or three heavy neutrinos to be
relatively light (e.g., at the keV mass scale). Such sterile
neutrinos are interesting in particle physics and cosmology. Note
that it is easier to satisfy the exact seesaw relation with an
appreciable value of $\varepsilon^{}_{\rm uv}$ by arranging $M^{}_i$
to lie in the keV, MeV or GeV range. This kind of low-scale seesaw
scenarios \cite{DeGouvea} might be technically natural, but they
have more or less lost the seesaw spirit. Of course, sufficiently
large $M^{}_i$ and sufficiently small $\theta^{}_{ik}$ can always
coexist to make the seesaw mechanism work in a natural way, but in
this traditional case the non-unitary effects are too small to have
any measurable consequences at low energies.

It is also worth pointing out that the seesaw-induced non-unitary
effects on $\mu^{}_{ij}$ and $\epsilon^{}_{ij}$ are rather different
from the case of making a naive assumption of the flavor mixing
between three active neutrinos and a few light sterile neutrinos
\cite{Sterile}. The latter can directly break the unitarity of the
$3\times 3$ active neutrino mixing matrix $V$ and then lift the GIM
suppression associated with $\mu^{}_{ij}$ and $\epsilon^{}_{ij}$.
This kind of non-unitary effects are not constrained by the seesaw
relation, and thus they are more arbitrary and less motivated from
the point of view of model building.

\vspace{0.4cm}

\framebox{\bf 4} ~ We have explored the seesaw-induced
non-unitary effects on the electromagnetic transition dipole moments
and radiative decays of three active neutrinos. We find that such
effects are possible to be comparable with or larger than the
standard (unitary) contributions, because the suppression induced by
the GIM mechanism in the latter case can be lifted. Our numerical
analysis has illustrated that the effective electromagnetic dipole moments
of three neutrinos and the rates of their radiative decays can be
maximally enhanced by a factor of ${\cal O}(10^2)$ and a factor of
${\cal O}(10^4)$, respectively, no matter whether the seesaw scale
is around or below the TeV energy scale. This observation
is new and nontrivial, and it clearly reveals an intrinsic and
presumably important correlation between the electromagnetic
properties of massive neutrinos and the origin of their small
masses. Such a correlation may even serve as a sensitive touch-stone
for the highly-regarded seesaw mechanism.

\vspace{0.4cm}

We thank W. Chao, Y.F. Li and S. Zhou for useful
discussions. This work was supported in part by the National Natural
Science Foundation of China under Grant No. 11135009.

\newpage

\begin{figure}[t]
  \begin{center}
  \includegraphics[width=1\textwidth]{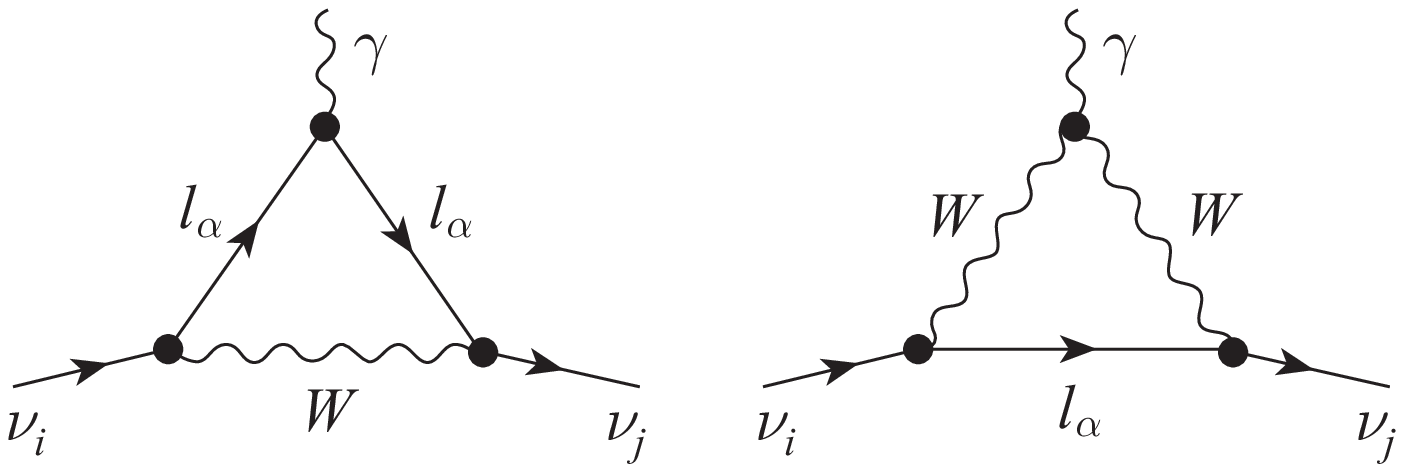} \\
  \end{center}
  \vspace{-0.4cm}
  \caption{The one-loop Feynman diagrams (and their charge-conjugate
  counterparts) contributing to the EMDMs of the Majorana neutrinos,
  where $\alpha = e, \mu, \tau$ and $i,j = 1,2,3$.}
\end{figure}
\begin{figure}[t]
  \begin{center}
  \includegraphics[width=1\textwidth]{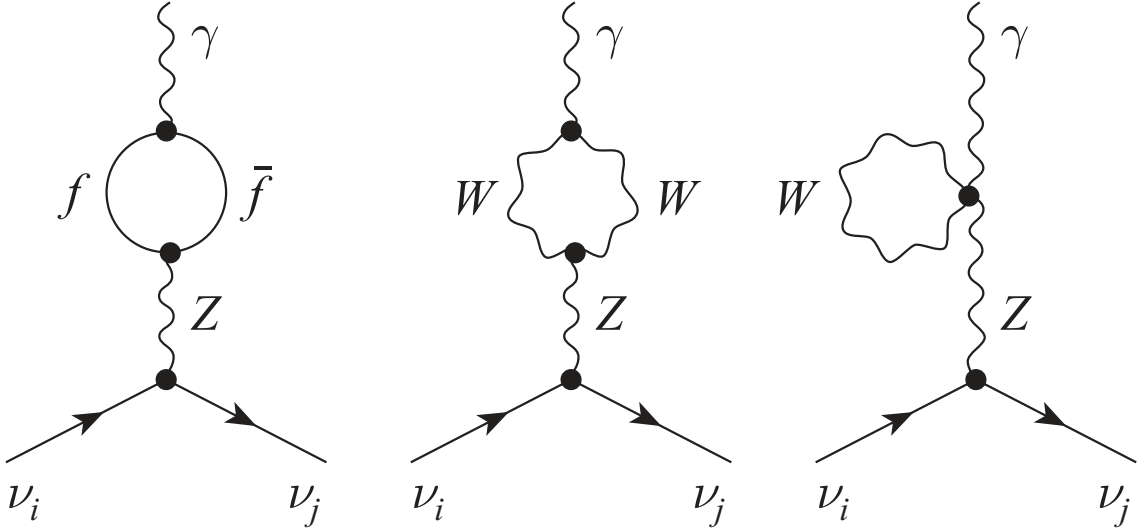} \\
  \end{center}
  \vspace{-0.4cm}
  \caption{The one-loop $\gamma$-$Z$ self-energy diagrams (and their
  charge-conjugate counterparts) associated with the EMDMs of massive Majorana
  neutrinos, where $f$ denotes all the SM fermions and $i,j = 1,2,3$.}
\end{figure}
\begin{figure}[t]
\vspace{-0.15cm}
  \begin{center}
  \includegraphics[width=0.97\textwidth]{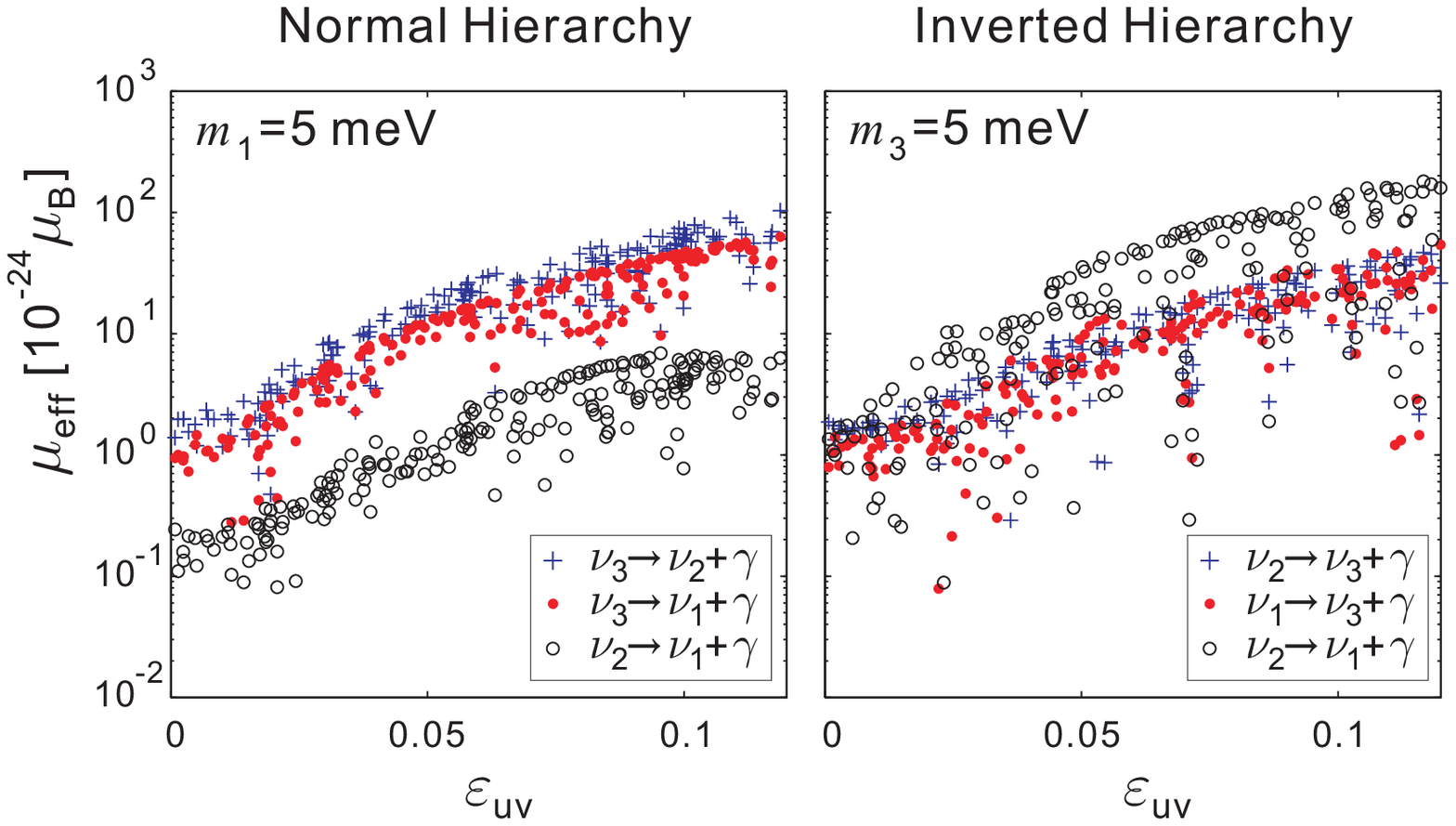} \\
  \end{center}
  \vspace{-0.4cm}
  \caption{Illustration of the seesaw-induced
  non-unitary effects on $\mu^{}_{\rm eff}$ for three active neutrinos.
  The standard (unitary) results correspond to $\varepsilon^{}_{\rm uv} =0$,
  and their uncertainties come from the three unknown CP-violating phases
  of $V^{}_0$.}
\end{figure}
\begin{figure}[t]
\vspace{-0.15cm}
  \begin{center}
  \includegraphics[width=0.97\textwidth]{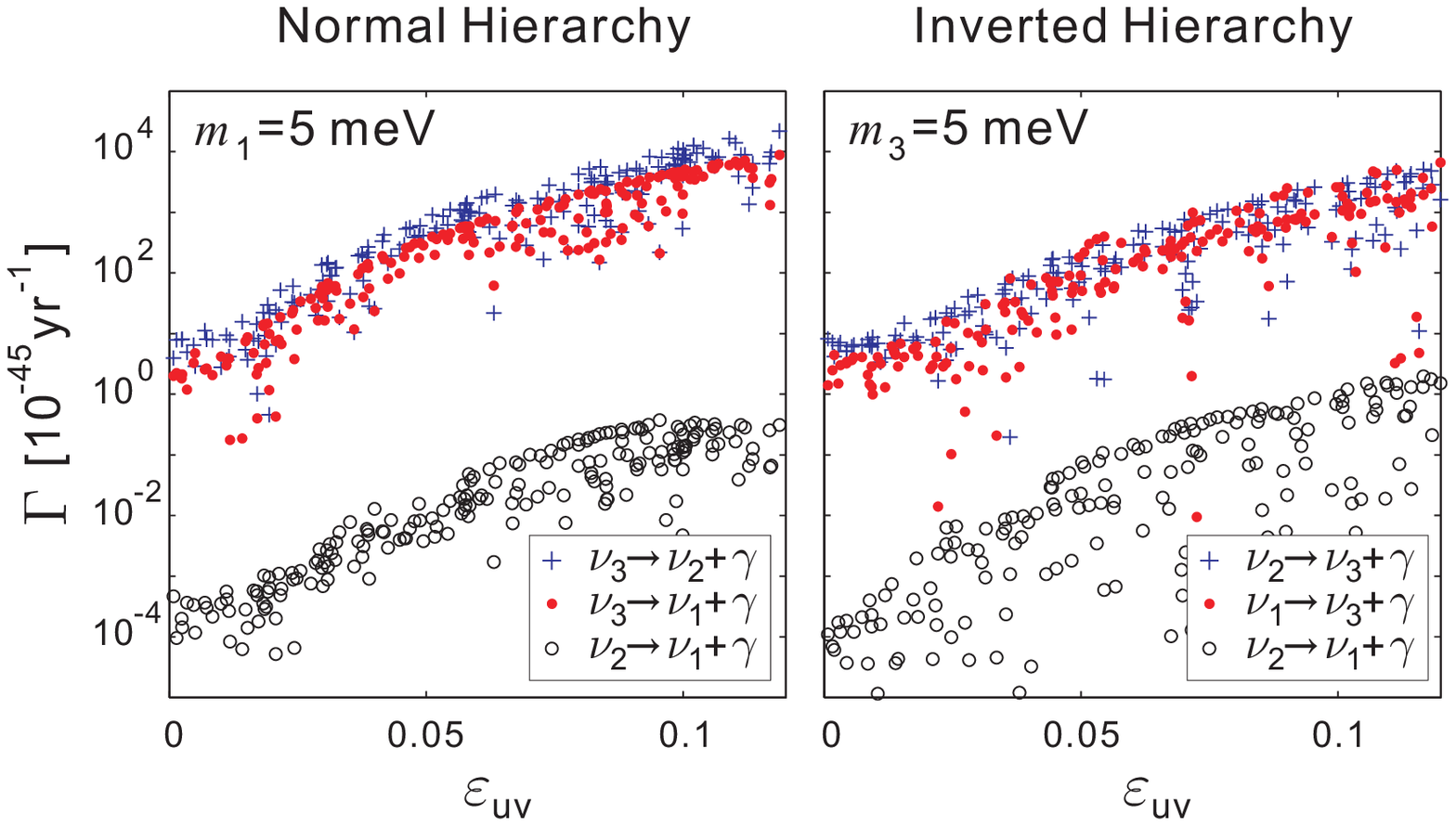} \\
  \end{center}
  \vspace{-0.4cm}
  \caption{Illustration of the seesaw-induced non-unitary
  effects on $\Gamma^{}_{\nu^{}_i \to \nu^{}_j + \gamma}$ for three active
  neutrinos. The standard (unitary) results correspond to
  $\varepsilon^{}_{\rm uv} =0$, and their uncertainties come from the three
  unknown CP-violating phases of $V^{}_0$.}
\end{figure}

\end{document}